# Thermopower analysis of the electronic structure around metal-insulator transition in $V_{1-x}W_xO_2$

Takayoshi Katase, Kenji Endo, and Hiromichi Ohta[*]

*Research Institute for Electronic Science, Hokkaido University, N20W10, Sapporo 001-0020, Japan*

[*]Corresponding author. E-mail: hiromichi.ohta@es.hokudai.ac.jp

Electronic structure across the metal-insulator (MI) transition of electron-doped $V_{1-x}W_xO_2$ epitaxial films ($x = 0$–$0.06$) grown on $\alpha$-$Al_2O_3$ substrates was studied by means of thermopower ($S$) measurements. Significant increase of $|S|$-values accompanied by MI transition was observed, and the transition temperatures of $S$ ($T_S$) decreased with $x$ in good linear relation with MI transition temperatures. $|S|$ values of $V_{1-x}W_xO_2$ films at $T > T_S$ were constant at low values of 23 $\mu$V K$^{-1}$ independently of $x$, which reflects a metallic electronic structure, whereas, those at $T < T_S$ almost linearly decreased with logarithmic W-concentrations. The gradient of $-213$ $\mu$V K$^{-1}$ agrees well with $-k_B/e \cdot \ln 10$ ($-198$ $\mu$V K$^{-1}$), suggesting that $V_{1-x}W_xO_2$ films have insulating electronic structures with a parabolic density of state around the conduction band bottom.

Vanadium dioxide ($VO_2$) has attracted considerable attention due to its ability to reversibly transform from a low-temperature insulator into a high-temperature metal at ~340 K.[1] The metal-insulator (MI) transition is accompanied by a structural change from monoclinic to tetragonal-type rutile structure. The transformation of crystal structure originates from dimerization of vanadium ions with accompanying the position shifting from linear chains along *c*-axis of rutile phase to zigzag type, resulting in a monoclinic structure. The structural change causes reconstruction of electronic structures to open up a charge gap of ~0.6 eV that abruptly changes both the electrical resistivity and infrared transmission.[2] These features of the MI transition for $VO_2$ appear promising for potential applications to electrical and optical switching devices, operating at room temperature (RT). Recently, reversible alternation of electronic properties from insulator to metal state was demonstrated by both electrostatic charge-doping[3] and hydrogenation,[4] which enables on-demand-tunable devices using the MI transition of $VO_2$.

However, driving mechanism of the MI transition in $VO_2$ is still not fully understood, i.e., it has been debated that the MI transition should be regarded as whether a structurally-driven Peierls transition with electron-phonon interaction or a Mott transition with strong electron-electron correlation.[5] Thus, intensive efforts have been devoted to experimentally observe electronic structure change of $VO_2$ across the MI transition mainly by spectroscopic techniques, such as X-ray photoemission spectroscopy (PES)[6] and angle resolved PES[7] for valence band structure observation, as well as X-ray absorption spectroscopy[8] for the conduction band structure, but the mechanism of the MI transition is still unclear. Further investigation on the electronic-structure evolution by another experimental means is inevitable for the elucidation of MI transition, which should give crucial information for fundamental physics as well as for practical device application of VO₂.

Here, we focused on thermopower ($S$) as a physical property to investigate the electronic structure across the MI transition, because $S$-values should be sensitive to significant changes in the electronic structure of VO₂ at $T_{MI}$. In general, $S$-value of metals (degenerate semiconductors) is basically expressed as $S = \frac{\pi^2}{3}\frac{k_B^2 T}{e}\left\{\frac{d\left[\ln(\sigma(E))\right]}{dE}\right\}_{E=EF}$ in Mott's equation, where $\sigma(E)$ is energy-dependent conductivity and $k_B$ is Boltzmann constant.[9] Meanwhile, that of semiconductors is expressed as $S = \frac{k_B}{e}(\frac{E_F - E_c}{k_B T} + A)$, assuming that only electrons contribute to the $S$-values and $E_F$ lies near the conduction band edge ($E_c$).[10] The $S$ can be simplified to $S = \frac{k_B}{e}(\ln\frac{N_c}{n_e} + A)$, where $N_c$ is effective density of state (DOS) of the conduction band, $n_e$ is carrier concentration, and $A$ is a transport constant that depends on the dominant scattering mechanism. $S$-values of metals are typically small and keep constant by reflecting the energy differential of DOS around the Fermi energy ($E_F$), but those of the semiconductors drastically change, depending on $n_e$, by reflecting the shape of DOS around conduction band bottom due to the $E_F$ shifts by carrier doping.[11] $S$ measurements around MI transition of VO₂ can be expected as a powerful tool to experimentally investigate their electronic structure evolutions.

In this Rapid Communication, we systematically investigated the $S$-values of electron-doped $V_{1-x}W_xO_2$ epitaxial films with different doping levels. A few $S$ measurements of un-doped VO₂ have been reported,[12-15] but there has been no report on electron-doped VO₂. Chemical substitution of VO₂ with aliovalent ions of $W^{6+}$ is a classical way to effectively dope electrons[16] and reduce the $T_{MI}$.[17] Abrupt changes in the *S*-values accompanied by MI transition were observed for the all films and the transition temperature of *S* decreased with *x* in good linear relation with $T_{MI}$. We examined the electronic-structure changes of $V_{1-x}W_xO_2$ films across the MI transition by means of *S* measurements.

$V_{1-x}W_xO_2$ films were fabricated on $(11\bar{2}0)$ α-Al$_2$O$_3$ single-crystalline substrates by pulsed laser deposition. A KrF excimer laser (wavelength of 248 nm, laser energy fluence of ~2 J cm$^{-2}$ pulse$^{-1}$, and repetition rate of 10 Hz) was used to ablate WO$_3$-added V$_2$O$_5$ polycrystalline target disks, which were prepared by sintering V$_2$O$_5$ and WO$_3$ powders mixed in a stoichiometric ratio of $V_2O_5 : WO_3 = (1-x)/2 : x$. The film composition of *x* was varied with the nominal composition of the targets. The growth temperature was fixed at 500 °C and oxygen partial pressure ($P_{O2}$) was optimized at 2.0 Pa because the ratio of resistivity change across MI transition is extremely sensitive to $P_{O2}$ during thin film growth.[18] After the deposition, the films were cooled to RT under the same oxygen pressure. The film thickness was fixed at ~20 nm, which was characterized by X-ray reflectivity measurement.

The film structures, including the crystalline phase and the orientation of the crystallites, were examined by X-ray diffraction (XRD, anode radiation: monochromatic CuK$\alpha_1$) at RT. Figure 1(a) shows the out-of-plane XRD patterns of $V_{1-x}W_xO_2$ films with various doping levels (*x* = 0, 0.01, 0.022, 0.06). For an un-doped VO$_2$ film (*x* = 0), *h*00 (*h* = 2, 3, 4) diffraction peaks of monoclinic VO$_2$ (M) phase were observed along with intense peaks of α-Al$_2$O$_3$ substrate. The full width at the half maximum value of the out-of-plane rocking curve for 200 (M) was 0.2°. Although many rectangular shaped grains were observed in the topographic AFM image [Fig. 1(b)], VO$_2$ films were heteroepitaxially grown on α-Al$_2$O$_3$ substrates, which was confirmed by reflection high energy electron diffraction (RHEED) pattern [inset of (b)] and the in-plane XRD measurements (data not shown). The epitaxial relationship was (100)[010] $V_{1-x}W_xO_2$ (M) ∥ (11$\bar{2}$0)[0001]α-Al$_2$O$_3$ as illustrated in Fig. 1(c). As $x$ increased in $V_{1-x}W_xO_2$ films, the peak intensity of 300 (M) weakened and disappeared at $x = 0.022$. Since the double lattice spacing along $a$-axis of monoclinic phase originates from the formation of vanadium-ion dimer, the disappearance of 300(M) diffraction proves the transformation from monoclinic to rutile-type structure. For $V_{1-x}W_xO_2$ films with $x \geq 0.022$, $h0l$ ($h = l$) diffraction peaks of tetragonal VO$_2$ (T) were observed, indicating that the structural transition temperature decreased below RT. Figure 1(d) shows optical transmission spectra of $V_{1-x}W_xO_2$ films. Optical transparency in the infrared region drastically decreased at $x \geq 0.022$, which is consistent with the structural transformation from M- to T-phase at RT. All obtained $V_{1-x}W_xO_2$ films with $x$ up to 0.06 were confirmed to be epitaxially grown on α-Al$_2$O$_3$ substrates and the crystalline orientation kept unchanged, independently of $x$.

Then we measured temperature dependence of electrical resistivity ($\rho$–$T$) by means of d.c. four probe method with van der Pauw electrode configuration. Figure 2(a) shows the $\rho$–$T$ curves normalized by $\rho$ at 350 K for $V_{1-x}W_xO_2$ epitaxial films with $x = 0$–0.06. The arrows indicate the position of $T_{MI}$, which is defined as the peak position of the derivative curve, $\dfrac{d[\log \rho]}{dT}$. The $\rho$ of un-doped VO$_2$ film showed a sharp resistivity jump at $T_{MI}$ of 338 K, which is similar to 341 K of VO$_2$ bulks.[1] Generally, epitaxial strains imposed on VO$_2$ films by substrates have a significant effect on $T_{MI}$. Compared to VO$_2$ films grown on (001) TiO$_2$ substrates,[19] where $T_{MI}$ is depressed down to below 300 K without intentional doping, the VO$_2$ films on α-Al$_2$O$_3$ substrates are not subjected to an epitaxial strain effect, presumably because lattice relaxation of VO$_2$ occurs at the interface of the α-Al$_2$O$_3$ substrate due to the difference in crystallographic symmetry. With increase of $x$, the $T_{MI}$ shifted to a lower temperature and became below RT at $x \geq$ 0.022, which is consistent with the decrease in the structural transition temperature observed in the XRD measurements.

The $S$-values were measured by giving a temperature difference of ~2 K in the film along the [010] direction, where the actual temperatures of both sides of V$_{1−x}$W$_x$O$_2$ film surface were monitored by two tiny thermocouples. The thermo-electromotive force ($\Delta V$) and $\Delta T$ were simultaneously measured, and the $S$-values were obtained from the slope of $\Delta V$–$\Delta T$ plots [inset of Fig. 2(b)], which ensures a linear relationship between $\Delta V$ and $\Delta T$. Figure 2(b) summarizes the $S$−$T$ curves. The obtained $S$-values were negative in the entire temperature range, indicating that n-type carriers are dominant in both the metal and insulating phases of V$_{1−x}$W$_x$O$_2$ films. As the temperature decreases, significant increase and the saturation of $S$-values were observed for all films. It should be noted that it was hard to measure $S$-values of un-doped VO$_2$ films at low temperature because of the high contact resistance > 1 MΩ, i.e. reliable thermoelectromotive force was not obtained at low temperature. The saturated $S$-values of V$_{1−x}$W$_x$O$_2$ films ($x =$ 0.01−0.06) decreased linearly with decrease of temperature down to zero, which is consistent with the linear decrease of $S$-$T$ for insulating phase of un-doped VO$_2$ bulk,[13] and suggesting that they are degenerate semiconductors. The transition temperatures ($T_S$), where $S$-values start to increase, are indexed by arrows. We compare the $x$ dependences of $T_S$ and $T_{MI}$ [Fig. 3(a)] extracted from $\rho$–$T$ [Fig. 2(a)] and $S$−$T$ [Fig. 2(b)]. $T_S$ and $T_{MI}$ were observed at almost the same temperature and monotonically decreased with an increase of $x$, which clearly indicate that the transition observed in $S$−$T$ originates from electronic structure reconstruction at $T_{MI}$.

For metallic phase at $T > T_S$, the $S$-values of $V_{1-x}W_xO_2$ films were constant at $-23$ $\mu V\ K^{-1}$ regardless of $x$ [Fig. 2(b)], which agrees well with the previously reported $S$-values of $\sim -20\ \mu V\ K^{-1}$ for the metallic phase of un-doped $VO_2$ bulks,[12,13] microbeams,[14] and films.[15] On the other hand, for the insulating phase at $T < T_S$, the saturated maximum $|S|$-values ($|S_{max}|$), which are defined as the $|S|$-values for intrinsic insulating phases,[15] steeply decreased from 205 $\mu V\ K^{-1}$ ($x = 0.01$) to 43 $\mu V\ K^{-1}$ as $x$ increased up to 0.06 [Fig. 2(b)]. $|S|$-values of insulating $V_{1-x}W_xO_2$ at low temperature showed $T$-linear tendency, suggesting that the $|S|$-value obeys Mott formula, $S = \dfrac{\pi^2}{3}\dfrac{k_B^2\,T}{e}\left\{\dfrac{d\left[\ln(\sigma(E))\right]}{dE}\right\}_{E=EF}$.[9] Therefore, we used Mott formula divided by $T$, $|S_{max}|$ / $T_{max}$, to compare the $|S_{max}|$-values of insulating $V_{1-x}W_xO_2$ films with different $x$ at the same temperature. As shown in Fig. 3 (b), $|S_{max}|$ / $T_{max}$ monotonically decreased with increasing $x$, suggesting that the $\left[\dfrac{\partial \mathrm{DOS}(E)}{\partial E}\right]_{E=EF}$ becomes moderate with an increase of $x$.

In order to construct the carrier density ($n_e$) dependence of $S$-values for the $V_{1-x}W_xO_2$ films, Hall effect measurement with van der Pauw electrode configuration was performed at RT, but reliable Hall voltages were not obtained, presumably due to the low carrier mobility ($\leq 0.1\ cm^2\ V^{-1}\ s^{-1}$) and high carrier concentration of the $V_{1-x}W_xO_2$ films.[20] Therefore, we used the W-concentration instead of $n_e$ from doping levels ($x$) in $V_{1-x}W_xO_2$ films and plotted $|S_{max}|$ at 300 K [Fig.4]. In general, semiconductors possessing a parabolic DOS show a linear relationship between $|S|$ and the log of carrier density ($\log n_e$)[21]: $|S| = -k_B/e \cdot \ln 10\ (\log n_e + C)$, where $C$ is parameter that depend on the types of materials. $|S_{max}|_{300K}$ almost linearly decreased from 266 down to 105 $\mu V\ K^{-1}$ as a function of log [W] with a gradient of $-213\ \mu V\ K^{-1}$ decade$^{-1}$, which agrees well with $-k_B/e \cdot \ln 10$ (= $-198$ μV K$^{-1}$ decade$^{-1}$). The linear decrease of $S$ against log [W] suggests that $S$-values of $V_{1-x}W_xO_2$ films at $T < T_s$ reflects insulating electronic structures with a parabolic DOS around the conduction band bottom in the doping range of $x = 0.01-0.06$, which is consistent with the calculated band structure of insulating $VO_2$.[22]

Here, we summarize the present results, comparing with the suggested electronic structure of $VO_2$.[2] In principal, lower-energy $t_{2g}$ state of the V 3$d$ orbital splits into $d_{\parallel}$ band and $\pi^*$ band. In the metallic T-phase, $d_{\parallel}$ band overlaps the $\pi^*$ band, and $E_F$ is located at the partially filled hybridized-band between the $d_{\parallel}$ and $\pi^*$ states. This scenario is consistent with the constant $S$-values of $-23$ μV K$^{-1}$ for $V_{1-x}W_xO_2$ films at $T > T_{MI}$, independently of W-concentration [Fig.4]. In the insulating M-phase, dimerization of V ions raises the $\pi^*$ band above $E_F$ and the $d_{\parallel}$ band splits into bonding- and antibonding-$d_{\parallel}$ states, creating a charge gap between $\pi^*$ band and antibonding-$d_{\parallel}$ band. Therefore, the steep decrease in the $|S_{max}|$-values with $x$ and the linear relation of $|S_{max}|$ against log [W], observed in $V_{1-x}W_xO_2$ films at $T < T_{MI}$, indicate that the doped carriers are simply accomodated in the $\pi^*$ band possessing a parabolic DOS in the doping rage of $x = 0.01-0.06$.

In summary, we investigated the $S$-values of electron-doped $V_{1-x}W_xO_2$ epitaxial films grown on α-$Al_2O_3$ substrates to experimentally examine the electronic-structure change across the MI transition. $/S/$ values of $V_{1-x}W_xO_2$ films at $T > T_S$ were independent of $x$ and remain constant at low values of 23 μV K$^{-1}$, which reflects the metallic electronic structure. On the other hand, those at $T < T_S$ almost linearly decreased with logarithmic W-concentrations. The gradient of $-213$ μV K$^{-1}$ agrees well with $-k_B/e \cdot \ln 10$ ($-198$ μV K$^{-1}$), suggesting that they have insulating electronic structures with a parabolic density of state around the conduction band bottom in the doping range of $x = 0.01-0.06$. The present results should provide crucial information not only for fundamental physics but also for practical device applications of $VO_2$.

## Acknowledgments

This work was supported by Grant-in-Aid for Scientific Research A (No. 25246023), Grant-in-Aid for Scientific Research on Innovative Areas (No. 25106007) from the Japan Society for the Promotion of Science (JSPS), and the Asahi Glass Foundation.

# References

[1] F. J. Morin, Phys. Rev. Lett. **3**, 34 (1959).

[2] J. B. Goodenough, J. Solid State Chem. **3**, 26 (1971).

[3] M. Nakano, K. Shibuya, D. Okuyama, T. Hatano, S. Ono, M. Kawasaki, Y. Iwasa, and Y. Tokura, Nature **487**, 459 (2012).

[4] J. Wei, H. Ji, W. Guo, A. H. Nevidomskyy, and D. Natelson, Nature Nanotechnol. **7**, 357 (2012).

[5] R. M. Wentzcovitch, W. W. Schulz, and P. B. Allen, Phys. Rev. Lett. **72**, 3389 (1994); A. Zylbersztejn and N. F. Mott, Phys. Rev. B **11**, 4383 (1975).

[6] E. Sakai, K. Yoshimatsu, K. Shibuya, H. Kumigashira, E. Ikenaga, M. Kawasaki, Y. Tokura, and M. Oshima, Phys. Rev. B **84**, 195132 (2011).

[7] K. Saeki, T. Wakita, Y. Muraoka, M. Hirai, T. Yokoya, R. Eguchi, and S. Shin, Phys. Rev. B **80**, 125406 (2009).

[8] M. Abbate, F. M. F. de Groot, J. C. Fuggle, Y. J. Ma, C. T. Chen, F. Sette, A. Fujimori, Y. Ueda, and K. Kosuge, Phys. Rev. B **43**, 7263 (1991).

[9] M. Jonson and G. D. Mahan, Phys. Rev. B **21**, 4223 (1980).

[10] H. Fritzsche, Solid State Commun. **9**, 1813 (1971).

[11] H. Ohta, Y. Masuoka, R. Asahi, T. Kato, Y. Ikuhara, K. Nomura, and H. Hosono, Appl. Phys. Lett. **95**, 113505 (2009); Y. Nagao, A. Yoshikawa, K. Koumoto, T. Kato, Y. Ikuhara, and H. Ohta, Appl. Phys. Lett. **97**, 172112 (2010).

[12] C. N. Berglund and H. J. Guggenheim, Phys. Rev. **185**, 1022 (1969).

[13] C. Wu, F. Feng, J. Feng, J. Dai, L. Peng, J. Zhao, J. Yang, C. Si, Z. Wu, and Y. Xie, J. Am. Chem. Soc. **133**, 13798 (2011).

[14] J. Cao, W. Fan, H. Zheng, and J. Wu, Nano Lett. **9**, 4001 (2009).

[15]D. Fu, K. Liu, T. Tao, K. Lo, C. Cheng, B. Liu, R. Zhang, H. A. Bechtel, and J. Wu, J. Appl. Phys. **113**, 043707 (2013).

[16]C. Tang, P. Georgopoulos, M. E. Fine, J. B. Cohen, M. Nygren, G. S. Knapp, and A. Aldred, Phys. Rev. B **31**, 1000 (1985).

[17]K. Shibuya, M. Kawasaki, and Y. Tokura, Appl. Phys. Lett. **96**, 022102 (2010).

[18]H. Kim, N. Charipar, M. Osofsky, S. B. Qadri, and A. Piqué, Appl. Phys. Lett. 104, 081913 (2014).

[19]Y. Muraoka and Z. Hiroi, Appl. Phys. Lett. **80**, 583 (2002).

[20]D. Ruzmetov, D. Heiman, B. B. Claflin, V. Narayanamurti, and S. Ramanathan, Phys. Rev. B 79, 153107 (2009).

[21]G. H. Jonker, Philips Res. Rep. **23**, 131 (1968).

[22]E. Caruthers and L. Kleinman, Phys. Rev. B **7**, 3760 (1973).

.

**Figure captions**

FIG. 1. (Color online) (a)–(c) Crystallographic characterization at RT for $V_{1-x}W_xO_2$ epitaxial films with $x$ = 0–0.06 grown on $(11\overline{2}0)\,\alpha\text{-}Al_2O_3$ substrates. (a) Out-of-plane XRD patterns. Crystalline phases and diffraction indices are noted above the corresponding diffraction peaks. M and T signify monoclinic and tetragonal structures, respectively. (b) Topographic AFM image of $VO_2$ film ($x$ = 0). Inset shows the RHEED pattern (azimuth [010]). (c) Schematic epitaxial relation of $V_{1-x}W_xO_2$ / $(11\overline{2}0)\,\alpha\text{-}Al_2O_3$. (d) Optical transmission spectra of the $V_{1-x}W_xO_2$ epitaxial films at RT. Optical transparency in the infrared region drastically decreases at $x \geq 0.022$.

FIG. 2. (Color online) Temperature dependences of the electrical resistivity ($\rho$) and the thermopower ($S$) of $V_{1-x}W_xO_2$ epitaxial films with $x$ = 0–0.06. (a) $\rho$–$T$ curves normalized by $\rho$ at 350 K, $\rho\,/\,\rho_{350K}$. Transition temperatures of $T_{MI}$, indicated by arrows, gradually decrease as $x$ increases. (b) $S$–$T$ curves. Inset shows the $\Delta V$ vs. $\Delta T$ plots measured at RT. Arrows denote the transition temperatures ($T_S$), where $S$-values start to increase.

FIG. 3. (Color online) (a) $x$ dependences of the transition temperatures of $T_{MI}$ and $T_s$, which are extracted from $\rho$–$T$ and $S$–$T$ curves in Figs. 2(a) and (b), respectively. Open and closed symbols represent $T_{MI}$ and $T_s$, respectively. For the metallic phase (upper

part), $S$-values of $V_{1-x}W_xO_2$ films remain constant at $-23$ μV K$^{-1}$ and independent of $x$.

(b) $x$ dependence of $|S_{max}|$ divided by $T_{max}$, which corresponds to

$$\frac{\pi^2}{3}\frac{k_B^2}{e}\left\{\frac{d\left[\ln(\sigma(E))\right]}{dE}\right\}_{E=EF}$$ . The $|S_{max}| / T_{max}$ value decreases with an increase of $x$.

FIG. 4. (Color online) Relationship between $|S_{max}|$ at 300 K and W-concentration for metallic T-phase ($T > T_{MI}$) and insulating M-phase ($T < T_{MI}$) of $V_{1-x}W_xO_2$ films. For $|S_{max}|$ values of M-phase, gradient of $-213$ μV K$^{-1}$ decade$^{-1}$ agrees well with $-k_B/e \cdot \ln 10$ ($= -198$ μV K$^{-1}$ decade$^{-1}$).

(a) 

(c) 

(b) 

(d) 

FIG. 1

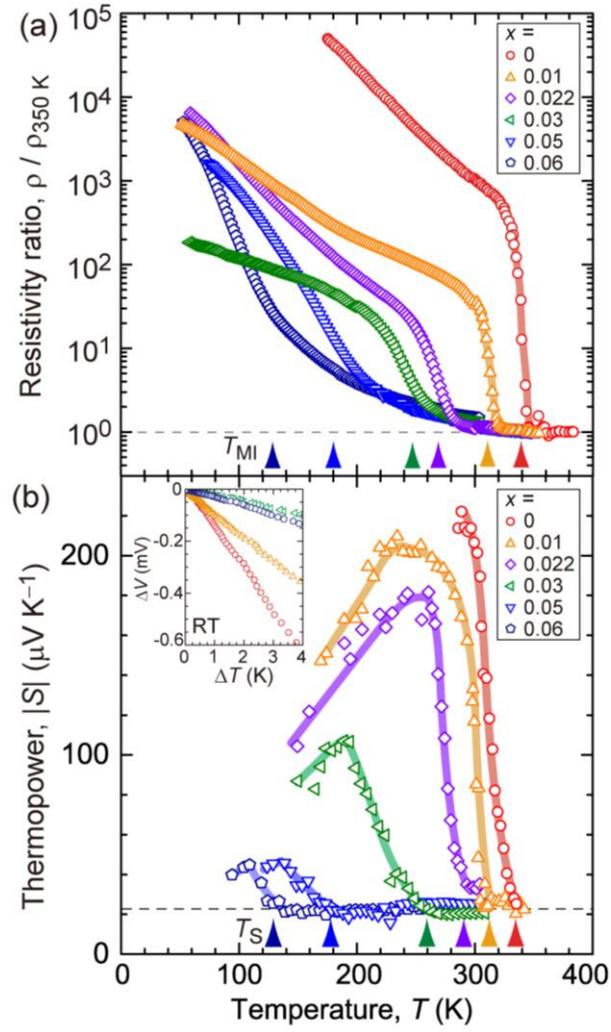

FIG. 2

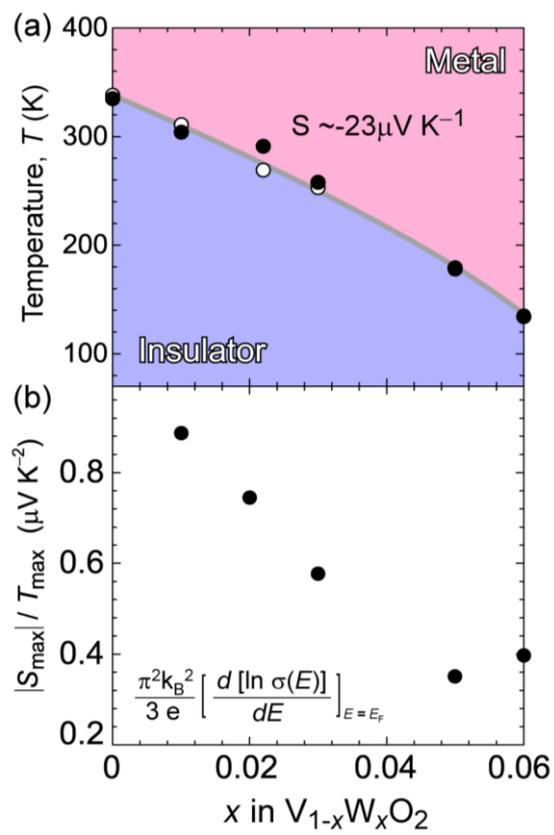

FIG. 3

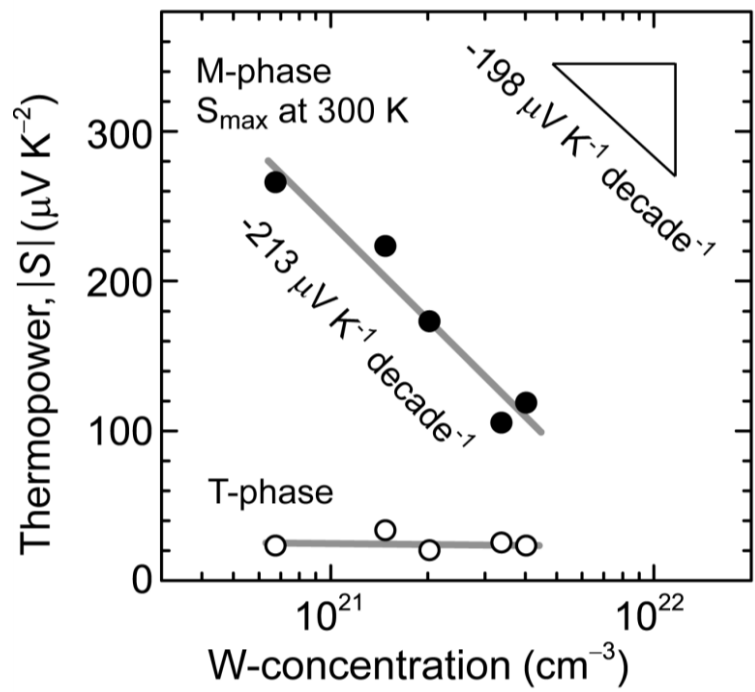

FIG. 4

18